# Observation of anyonic Bloch oscillations

Weixuan Zhang[1*], Hao Yuan[1*], Haiteng Wang[1], Fengxiao Di[1], Na Sun[1], Xingen Zheng[1], Houjun Sun[2], and Xiangdong Zhang[1$]

[1]Key Laboratory of advanced optoelectronic quantum architecture and measurements of Ministry of Education, Beijing Key Laboratory of Nanophotonics & Ultrafine Optoelectronic Systems, School of Physics, Beijing Institute of Technology, 100081, Beijing, China

[2] Beijing Key Laboratory of Millimeter wave and Terahertz Techniques, School of Information and Electronics, Beijing Institute of Technology, Beijing 100081, China

*These authors contributed equally to this work. $Author to whom any correspondence should be addressed. E-mail: zhangxd@bit.edu.cn

**Bloch oscillations are exotic phenomena describing the periodic motion of a wave packet subjected to the external force in a lattice, where the system possessing single- or multiple-particles could exhibit distinct oscillation behaviors. In particular, it has been pointed out that quantum statistics could dramatically affected the Bloch oscillation even in the absence of particle interactions, where the oscillation frequency of two pseudofermions with the anyonic statistical angle being $\pi$ becomes half of that for two bosons. However, these statistic-dependent Bloch oscillations have never been observed in experiments up to now. Here, we report the first experimental simulation of anyonic Bloch oscillations using electric circuits. By mapping eigenstates of two anyons to modes of designed circuit simulators, the Bloch oscillation of two bosons and two pseudofermions are verified by measuring the voltage dynamics. It is found that the oscillation period in the two-boson simulator is almost twice of that in the two-pseudofermion simulator, which is consistent with the theoretical prediction. Our proposal provides a flexible platform to investigate and visualize many interesting phenomena related to particle statistics, and could have potential applications in the field of the novelty signal control.**

Bloch oscillations (BOs) were originally proposed for electrons in crystals, which are characterized as the coherent oscillatory motion of electrons in a periodic potential driven by an external DC electric field [1, 2]. After a long-lasting debate about the correctness of BOs [3, 4], the effective Hamiltonians leading to BOs and their frequency-domain counterparts, called the Wannier-Stark ladder, were confirmed [5]. The first experimental observation of BOs is based on the semiconductor superlattices [6], and a few years later, atoms in an optical potential were also established to demonstrate such a novel effect [7-10]. In fact, the BO is a universal wave phenomenon, hence, it has also been observed in various classical wave systems [11-20], such as the coupled optical waveguides [11-19] and acoustic superlattices [20].

On the other hand, in the few-body quantum systems described by the Bose-Hubbard or Fermi-Hubbard model, many previous investigations have shown that BOs could be significantly modified by the strong particle interaction [20-29]. In particular, the frequency doubling of BOs for two strongly correlated Bosons, that is called fractional BOs, is experimentally observed based on a photonic lattice simulator [29], where the two-boson dynamics is directly mapped to the propagation of light fields in the designed two-dimensional waveguide array.

Except for bosons and fermions, anyons are quantum quasi-particles with the statistics intermediate between them [30-33]. Anyons play important roles in several areas of modern physics research, such as fractional quantum Hall systems [34-37] and spin liquids [38-40]. Another potential application of non-Abelian anyons is to realize the topological quantum computation [41]. Interestingly, a previous theoretical work has shown that two non-interacting anyons could exhibit exotic BOs, where the frequency halving of BOs for two pseudofermions exists if the ratio of the applied external force to the hopping rate is less than or equal to 0.5 [42]. While, the experimental observation of anyonic BOs is still a great challenging in condensed-matter systems, ultracold quantum gases and other classical wave systems. In this case, a newly accessible and fully controllable platform is expected to be constructed to simulate the anyonic BOs with novel behaviors.

In this work, we demonstrate both in theory and experiment that the anyonic BOs can be simulated by designed electric circuits. Using the exact mapping of two anyons in the external forcing to modes of designed circuit lattices, the periodic breathing dynamics of voltages have been observed by time-domain measurements in both two-boson and two-pseudofermion circuit simulators. In particular, we find that the oscillation frequency in the two-boson simulator is almost twice of that in the two-pseudofermion simulator, that is consistent with the theoretical prediction. Our work provides a flexible platform to implement many interesting phenomena depended on particle statistics, and could have potential applications in the field of the intergraded circuit design and the electronic signal control.

**Results**

**The theory of simulating anyonic Bloch oscillations by electric circuits.** We consider a pair of correlated anyons hopping on a one-dimensional (1D) chain subjected to an external force $F$. In this case, the system can be described by the extended version of the anyon-Hubbard model as:

$$H = -J\sum_{l=1}^{N}(a_l^+ a_{l+1} + a_{l+1}^+ a_l) + 0.5U\sum_{l=1}^{N} n_l(n_l - 1) + F\sum_{l=1}^{N} ln_l, \qquad (1)$$

where $a_l^+$ ($a_l$) and $n_l = a_l^+ a_l$ are the creation (annihilation) and particle number operators of the anyon at the *l*th lattice site, respectively. *N* is the number of lattice sites. *J* is the single-particle hopping rate between adjacent sites, and *U* defines the on-site interaction energy. The anyonic creation and annihilation operators obey the generalized commutation relations as:

$$a_k^+ a_l - a_l^+ a_k e^{i\theta sgn(l-k)} = \delta_{lk}, a_l a_k - a_k a_l e^{i\theta sgn(l-k)} = 0, \qquad (2)$$

where $\theta$ is the anyonic statistical angle, and $sgn(x)$ equals to -1, 0 and 1 for $x<0$, $x=0$ and $x>0$, respectively. It is worthy to note that anyons with $\theta = \pi$ are "pseudofermions" as they behave like fermions off site, while being bosons on site. The two-anyon solution can be expanded in the Fock space as:

$$|\psi> = \frac{1}{\sqrt{2}} \sum_{m,n=1}^{N} c_{mn} a_m^+ a_n^+ |0>, \qquad (3)$$

where $|0>$ is the vacuum state and $c_{mn}$ is the probability amplitude with one anyon at the site *m* and the other one at the site *n*. Under the restriction of anyonic statistics, the relationship of $c_{mn} = e^{i\theta sgn(n-m)} c_{nm}$ is satisfied. Substituting Eqs. (1) and (3) into the Schrödinger equation $H|\psi> = \varepsilon|\psi>$, we obtain the eigen-equation with respect to $c_{mn}$ as [42]:

$$\varepsilon c_{mn} = -J[e^{i\theta(\delta_{m,n}+\delta_{m+1,n})} c_{m(n-1)} + e^{-i\theta(\delta_{m,n}+\delta_{m-1,n})} c_{m(n+1)} + c_{(m-1)n} + c_{(m+1)n}]$$

$$+ U\delta_{mn} c_{mn} + F(m+n) c_{mn} . \qquad (4)$$

We note that Eq. (4) can also be regarded as the eigen-equation describing a single particle hopping on the 2D lattice with the spatially modulated on-site potential and hopping configuration, as shown in Fig. 1a with *U*=0. In this case, the probability amplitude for the 1D two-anyon model with one anyon at the site *m* and the other at the site *n* is directly mapped to the probability amplitude for the single particle locating at the site (*m*, *n*) of the 2D lattice. The position-dependent on-site potential could simulate the effect of external force. Moreover, the hopping of the single particle along a certain direction in the 2D lattice represents the hopping of one anyon in the 1D lattice. In this case, the behavior of two anyons in the 1D lattice can be effectively simulated by a single particle in the mapped 2D lattice, that inspires the design of classical simulators to study the statistic dependent anyonic physics.

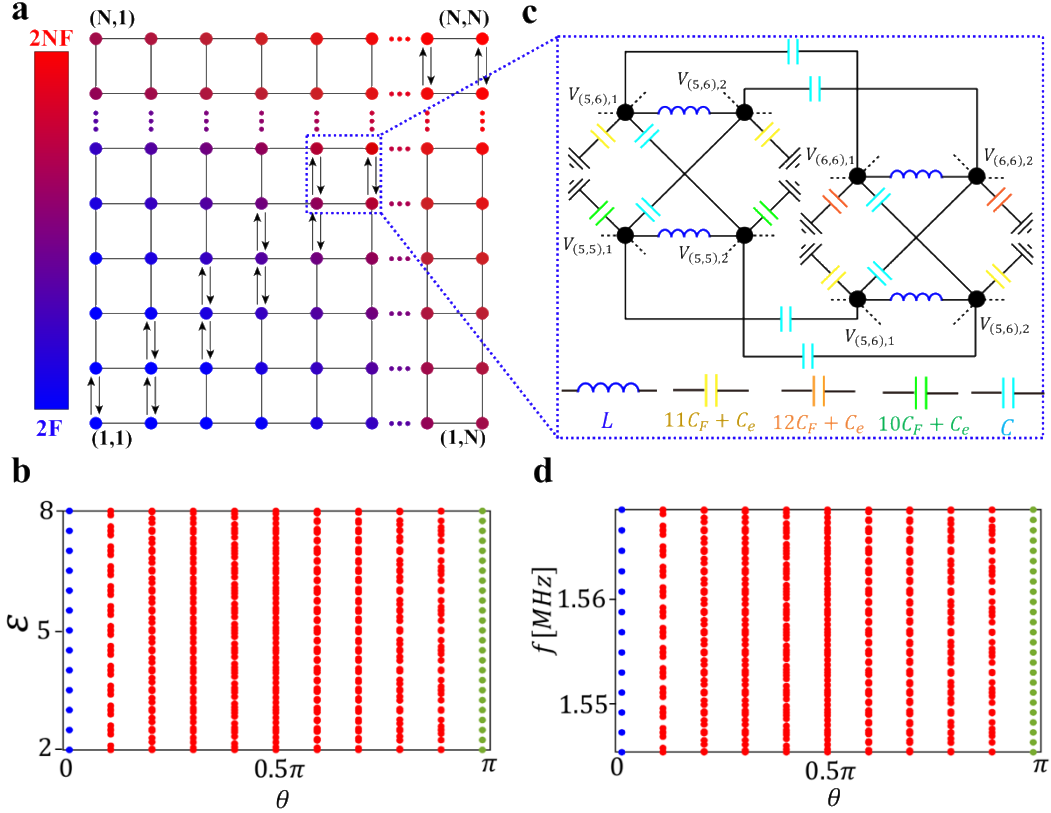

**Figure 1. The schematic diagram of designed circuit simulators for the Bloch oscillation of a pair of anyons.** (a). The mapped 2D lattice of the single particle for simulating the 1D two-anyon effect in the absence of on-site interaction under an external forcing. The color represents the value of on-site potential related to the external forcing. The arrow corresponds to the hopping rate with a complex phase $e^{\pm i\theta}$. (c). The schematic diagram for a part of designed circuit simulator with $\theta = \pi$ corresponding to four lattice sites enclosed by the blue dash block in Fig. 1a. (b) and (d) The calculated eigen-energies of two anyons and eigen-frequencies of the circuit simulator as a function of the statistical angle $\theta$.

One of the fascinating phenomena dominated by the quantum statistics in Eq. (4) is that the BO frequency of two pseudofermions ($\theta = \pi$) becomes half of that for two bosons ($\theta = 0$), when the on-site particle interaction is zero ($U=0$) and the ratio of the external forcing to the hopping rate satisfies $F/J \leq 0.5$ [42]. To clarify this effect, the evolution of two-anyon eigen-energies as a function of $\theta$ is displayed in Fig. 1b with $J=1$, $U=0$ and $F=0.5$. For the sake of clearness, the energies in the range of ($4F$, $16F$) are plotted. Moreover, to avoid the finite-size effect, only eigen-energies with their eigenmodes showing the largest overlap with the center of lattices are kept. It is seen that the eigen-spectrum is in the form of the Wannier-Stark ladder of two non-interacting bosons (blue dots for $\theta = 0$), where degenerated eigen-energies $\varepsilon_{mn} = (m+n)F$ are equally spaced by $\Delta\varepsilon = F$. When the statistical angle is non-zero, the associated eigen-energies are not equidistant any more, leading to the disappearance of the BOs. While, as for the case of pseudofermions ($\theta = \pi$), the eigen-spectrum (green dots) turns out to be nearly equal-spaced again with $\Delta\varepsilon = F/2$, which makes the BO frequency of two pseudofermions become half of that for two non-interacting bosons. Although the statistic-induced halving of the BO frequency is a very interesting phenomenon, up to now, the experimental observation of such an exotic effect is still an intractable challenge for both quantum platforms and classical simulators.

Based on the similarity between circuit Laplacian and lattice Hamiltonian [43-59], electric circuits can be used as an extremely flexible platform to fulfill the above mapped 2D lattice with different statistical angles. Fig. 1c illustrates the schematic diagram for a part of designed circuit simulator with $\theta = \pi$, that corresponds to four lattice sites enclosed by the blue dash block in Fig. 1a. Here, a pair of circuit nodes connected by the inductor $L$ are considered to form an effective site in the 2D lattice model. The voltages at these two nodes are marked by $V_{(m,n),1}$ and $V_{(m,n),2}$, which are suitably formulated to form a pair of pseudospins

$V_{\uparrow(m,n),\downarrow(m,n)} = (V_{(m,n),1} \pm V_{(m,n),2})/\sqrt{2}$. To simulate the real-valued hopping rate, two capacitors (the capacitance equals to $C$) are used to directly link adjacent nodes without a cross. As for the realization of hopping rate with a phase ($e^{\pm i\pi}$), two pairs of adjacent nodes are cross connected via $C$. And, the position-dependent capacitors $(m+n)C_F$ are used for grounding to simulate the spatially modulated on-site potential induced by the external forcing. Moreover, the extra capacitor $C_e$, that is crucial for the achievement of anyonic BOs in circuits demonstrated below, is also added to connect each circuit node to the ground. Through the appropriate setting of grounding and connecting, the circuit eigen-equation can be derived as:

$$(f_0^2/f^2 - 4 - C_e/C)V_{\downarrow,mn} = -e^{-i\pi(\delta_{m,n}+\delta_{m,n+1})}V_{\downarrow,m(n+1)} - e^{i\pi(\delta_{m,n}+\delta_{m+1,n})}V_{\downarrow,m(n-1)}$$
$$-V_{\downarrow,(m+1)n} - V_{\downarrow,(m-1)n} + (m+n)(C_F/C)V_{\downarrow,mn}, \quad (5)$$

where $f$ is the eigen-frequency ($f_0 = 1/2\pi\sqrt{CL/2}$) of the designed circuit, and $V_{\downarrow,(m,n)} = (V_{(m,n),1} - V_{(m,n),2})/\sqrt{2}$ represents the voltage of pseudospin at the circuit node $(m, n)$. Details for the derivation of circuit eigen-equations are provided in S1 of Supplementary Materials. It is shown that the eigen-equation of the designed electric circuit possesses the same form with Eq. (4). In particular, the probability amplitude for the 1D two-pseudofermion model $c_{mn}$ is mapped to the voltage of pseudospin $V_{\downarrow,(m,n)}$ at the circuit node $(m, n)$. The eigen-energy ($\varepsilon$) of two anyons is directly related to the eigen-frequency ($f$) of the circuit as $\varepsilon = f_0^2/f^2 - 4 - C_e/C$ with other parameters being $J = 1$ and $F = C_F/C$. It is worthy to note that the method for designing the circuit simulator is applicable to other statistical angles $\theta = \frac{v}{o}\pi$ ($v$ and $o$ are integers), where the complex hopping rate $Je^{\pm i\frac{v}{o}\pi}$ could be realized by suitably braiding the connection pattern of $o$ adjacent circuits nodes in a single lattice site [43, 44]. In this case, the relationship between $\varepsilon$ and $f$ with different values of $\theta$ stays the same.

To analyze the behavior of BOs in the circuit simulator with respect to $\theta$, eigenfrequencies of the designed circuit as a function of the statistical angle $\theta$ are calculated, as shown in Fig. 1d. Here, only the eigenfrequencies related to the eigen-energies in the range of ($4F$, $16F$) are plotted. The parameters are set as C=10$pF$, $C_e$=2$nF$, L=10$uH$ and $C_F$=5$pF$. It is noted that, due to the nonlinear relationship between the eigen-frequency of circuit simulator and the eigen-energy of two anyons ($f = f_0/(\varepsilon + 4 + C_e/C)^{1/2}$), the distribution of eigen-spectrum for the circuit simulator should deviate from that of two anyons with equal-spacings. In fact, such a deviation could be eliminated by setting an extremely large grounding capacitor $C_e$ (See S2 in the Supplementary Materials for details). In this case, we can see that the evolution of eigen-frequencies for the circuit simulator versus the statistical phase $\theta$ possesses the same trend with that of the two-anyon eigen-energy. In particular, the nearly equal-spaced eigen-spectrum of the circuit simulator with $\theta = 0$ ($\Delta f_B \approx 1.863 kHz$) exists, that is analog to the Wannier Stark ladder of a pair of non-interacting bosons. Additionally, the eigen-spectrum of the designed circuit for pseudofermions is also nearly equal-spaced with the spacing being $\Delta f_f \approx 0.931 kHz$, which is nearly the half of $\Delta f_B$. With such a good correspondence between the eigen-spectrum of designed 2D circuit and the 1D two-anyon model, the behavior of the quantum statistics dominated BO can be effectively simulated by the designed circuit simulator.

**Numerical results of simulating anyonic Bloch oscillations in electric circuits.** In this part, using the designed circuit simulator with $23 \times 23$ node pairs (corresponding to the 1D two-anyon model with *N*=23 lattice sites), we numerically simulate the behavior of BOs for two non-interacting anyons with the statistical angle being chosen as $\theta = 0$ and $\theta = \pi$, respectively. To observe such the statistics-dependent BO in our designed electric circuits, we perform the time-domain simulation of voltage dynamics using the LTSpice software. Here, the values of $C$, $C_F$, $C_e$ and $L$ are taken as $10pF$, $5pF$, $2nF$ and $10uH$ (the same to those used in Fig. 1d), respectively.

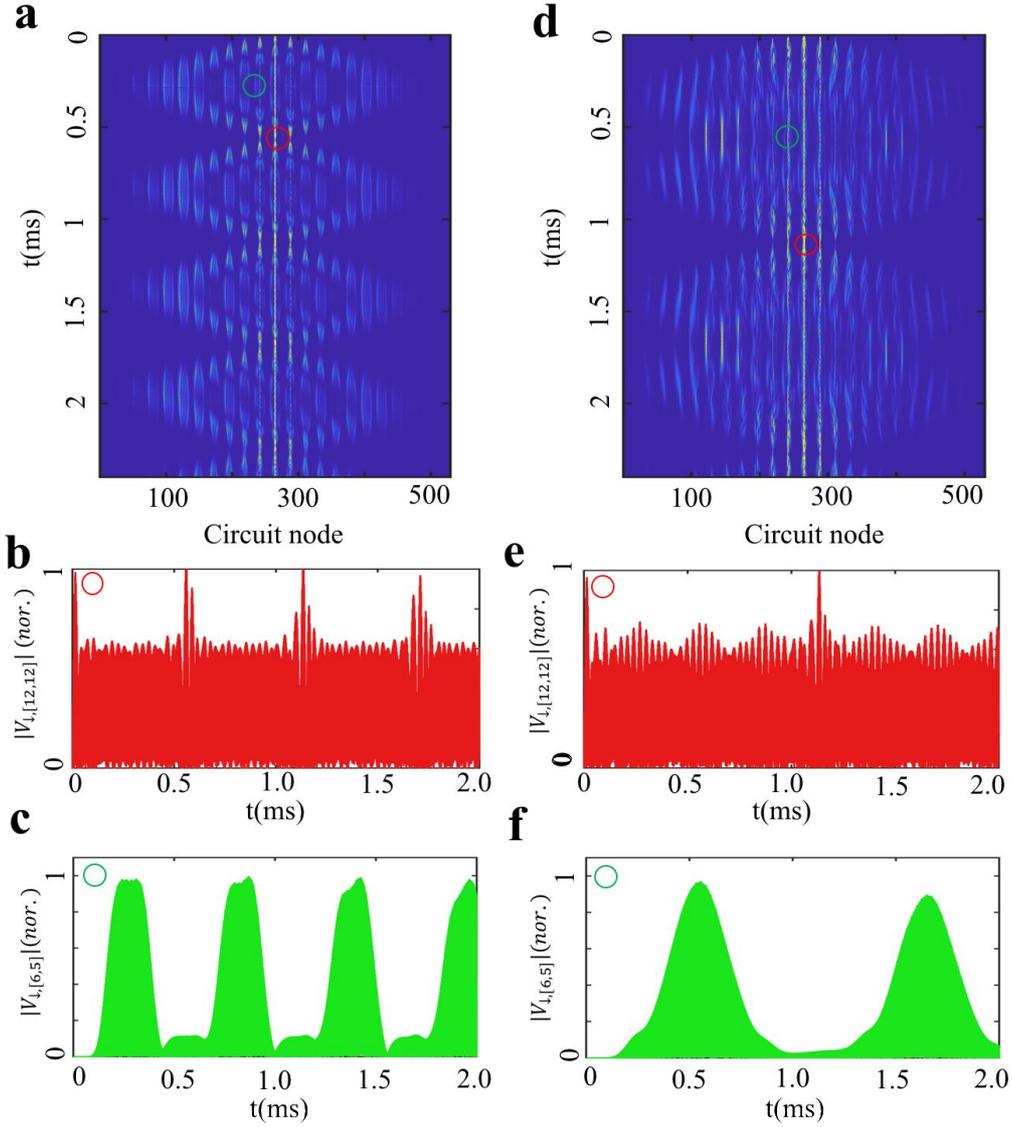

**Figure 2. Numerical results for simulating anyonic Bloch oscillations in electric circuits.** The time-dependent evolution of pseudospin $|V_{\downarrow,[m,n]}(t)|^2$ at each node in the 2D circuit simulator with $\theta = 0$ for (a), and $\theta = \pi$ for (d). The simulated voltage signals of $|V_{\downarrow,[12,12]}(t)|^2$ and $|V_{\downarrow,[6,5]}(t)|^2$ in the 2D circuit simulator with $\theta = 0$ for (b) and (c), and

$\theta = \pi$ for (e) and (f).

At first, we focus on the circuit simulator of two non-interacting bosons with $\theta = 0$. In this case, the excitation frequency is set as 1.56MHz, which locates in the region with equally spaced eigen-values at $\theta = 0$, as shown in Fig. 1d. The voltage-pseudospin could be suitably excited by setting the input signal with $V_{(12,12),1} = V_0$ and $V_{(12,12),2} = -V_0$ ($V_0 = 1V$). Fig. 2a displays the time-dependent evolution of pseudospin $|V_{\downarrow,[m,n]}(t)|^2$ at each node in the 2D circuit simulator. It is shown that the absolute value of pseudospin displays the periodic breathing dynamics, indicating the appearance of BOs. To further illustrate the simulated BOs of two non-interacting bosons, the time-dependent voltage signals $|V_{\downarrow,[m,n]}(t)|$ at circuit nodes (($m=12$, $n=12$) and ($m=6$, $n=5$)), which are marked by red and green circles in Fig. 2a, are calculated, and the results are presented in Figs. 2b and 2c. We can see that the revival of voltage-pseudospin is clearly verified. Additionally, it is noted that the associated oscillation period is about 0.545ms, which is nearly consistent with the period $T_B = \frac{1}{\Delta f_B} = 0.537ms$ calculated by the eigen-spectrum in Fig. 1d.

Next, we simulate the BO of two non-interacting pseudofermions by the designed electric circuit with $\theta = \pi$. The excitation frequency is also set as 1.56MHz, locating in the equally spaced region of eigen-frequencies at $\theta = \pi$. Similar to the two-boson case, the input signal is set as $[V_{1,(12,12)} = V_0, V_{2,(12,12)} = -V_0]$ to excite the voltage-pseudospin. The time-dependent evolution of $|V_{\downarrow,[m,n]}(t)|^2$ at each node in the 2D circuit simulator (with $\theta = \pi$) is shown in Fig. 2d. Moreover, the calculated voltage signals of $|V_{\downarrow,[12,12]}(t)|$ and $|V_{\downarrow,[6,5]}(t)|$ are presented in Figs. 2e and 2f. We can see that the periodic breathing dynamics of voltage could also appear

in the circuit simulator for two pseudofermions. And, the oscillation period is about 1.085ms, which is in a good consistence with $T_f = \frac{1}{\Delta f_f} = 1.074ms$ predicted by the eigen-spectrum. Moreover, comparing with the result of bosonic circuit simulators, we find that the oscillation period in the two-boson simulator is almost twice of that in the two-pseudofermion simulator. This phenomenon is consistent with the theoretical prediction based on the 1D anyon-Hubbard model (See S3 of Supplementary Materials for details).

It is worthy to note that above results only focus on the two-anyon model at a fixed excitation frequency 1.56MHz and constant values of $C_F = 5pF$ and $C_e = 2nF$. In S4 of Supplementary Materials, we also simulate the anyonic BOs by our designed electric circuits with different excitation frequencies, external forces and grounding capacitor $C_e$. It is shown that the smaller the external force is, the larger the oscillation period and amplitude become. Moreover, we find that more ideal BOs could be realized with a larger value of $C_e$, that could make the eigen-spectrum of circuit simulators become equally spaced.

**Experimental observation of anyonic Bloch oscillations in electric circuits.** To experimentally observe the anyonic BOs, the designed circuit simulator is fabricated, where the corresponding parameters are the same to that used in simulations. The photograph image of the circuit sample is presented in Fig. 3a and the enlarged views of front and back sides are plotted in right insets. Here, a single printed circuit board (PCB), which contains $23 \times 23$ node pairs, is applied for the circuit. It is noted that our fabricated circuit simulator could perform the BO of two anyons with $\theta = 0$ ($\theta = \pi$) when the switches (enclosed by white blocks) located around the diagonal line of the sample are opened (closed). This is due to the fact that these switches could change the connection pattern between adjacent circuit nodes around the diagonal line from direct connections to cross connections. In this case, if two pairs of adjacent circuit nodes are directly (cross) connected through the capacitor $C$ (framed by red circles), the hopping rate without (with) a phase ($e^{\pm i\pi}$) could be realized, that corresponds to the two-boson (two-pseudofermion) circuit simulator. Moreover, the position-dependent grounding capacitors $(m+n)C_F$ (framed by blue circles) are used to simulate the external forcing. The inductor $L$ and capacitor $C_e$ are enclosed by the pink and green frames. Additionally, the tolerance of the circuit elements is only 1% to avoid the detuning of circuit responses. Details of the sample fabrication is provided in Methods.

Firstly, we measure the temporal dynamics of the fabricated electric circuits with $\theta = 0$ (open switches), where two circuit nodes are excited by $[V_{(12,12),1} = V_0, V_{(12,12),2} = -V_0]$, and the associated frequency is 1.56MHz. Details of the experimental measurements are provided in Methods. Fig. 3c displays the measured envelope curve of voltage signal $|V_{\downarrow,[6,5]}(t)|$ in the time-domain. We note that the damped oscillation period is about 0.54ms, which is nearly consistent with the simulation result. The decay of the revival voltage is due to the large lossy effect, resulting from the resistive loss of linking wires and the low Q-factor of the applied inductor.

Then, we turn to the fabricated circuit with $\theta = \pi$ (close switches), and the excitation signal is the same to that used for bosonic circuit simulators. As shown in Fig. 3d, we measure the time-dependent evolution of $|V_{\downarrow,[6,5]}(t)|$. It is shown that the damped oscillation period is about 1.08ms, which is also in a good consistence with the simulated result. Comparing to the measured period of circuit simulator with $\theta = 0$, we note that the Bloch oscillation frequency related to a pair of pseudofermions ($\theta = \pi$) is half of that for two bosons ($\theta = 0$), being consistent with the theoretical prediction. Similar to the bosonic results, the significant decay of the voltage signal is also resulting from the large lossy effect in the fabricated circuit.

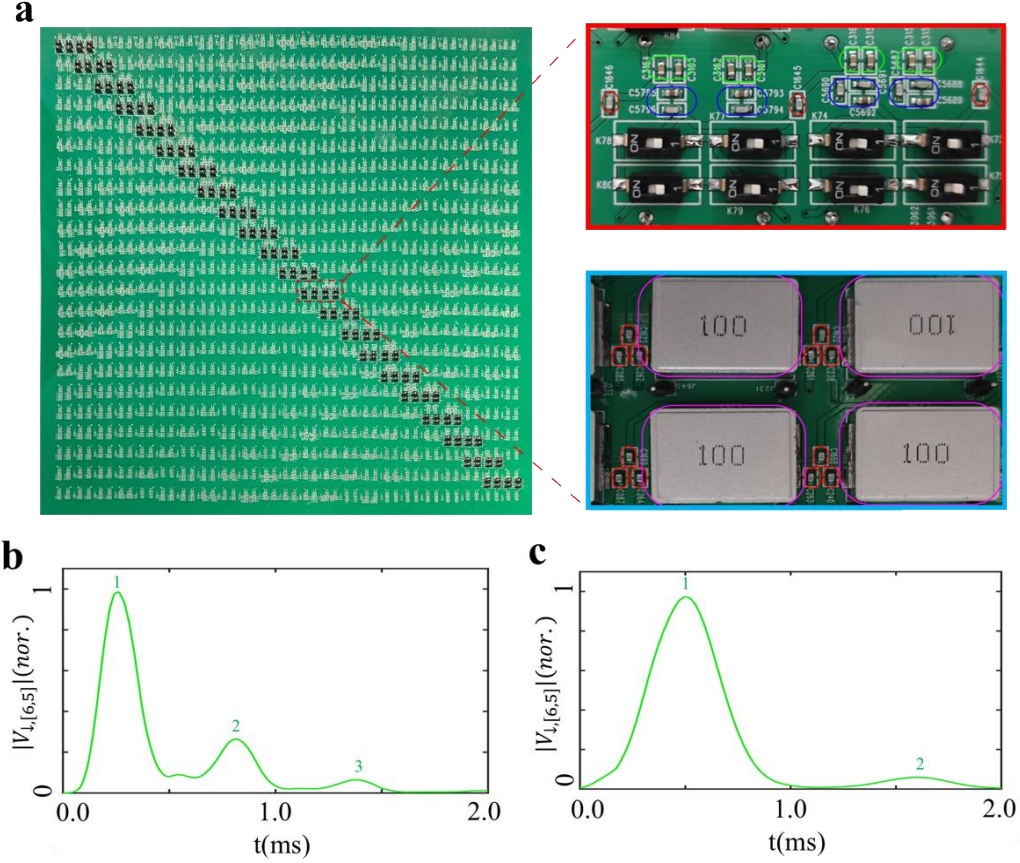

**Figure 3. Experimental results for observing anyonic Bloch oscillations.** (a) and (b). The photograph images of the fabricated circuit simulator. The enlarged views for the front and back sides are shown in bottom insets. (c) and (d) The measured envelope curve of $|V_{\downarrow,[6,5]}(t)|$ in the time-domain for the fabricated circuit simulators with ($\theta = 0$) and ($\theta = \pi$).

**Discussion and conclusion.**

With the advantage of diversity and flexibility for circuit elements, except for the above designed $LC$ circuit with matched stationary eigen-equations to the 1D two-anyon system, we can design another kind of electric circuits, which is based on resistances and capacitances, to precisely match the time-dependent Schrödinger equation of two pseudofermions and two bosons. In this case, the BOs dominated by the quantum statistics can also be observed in the designed *RC* circuit. The detailed results are given in S5 of Supplementary Materials.

In conclusion, we have experimentally demonstrated that electric circuits can be used as a flexible simulator to investigate the statistics-dominated BO, where the oscillation frequency of two pseudofermions is half of that for two bosons in the absence of on-site interaction. Using the exact mapping of two anyons in the external forcing to modes of designed circuit lattices, the periodic breathing dynamics of voltage in circuit simulators for $\theta = 0$ and $\theta = \pi$ have been observed. Our proposal provides a flexible simulator to investigate and visualize many interesting phenomena related to the particle statistics, and could have potential applications in the field of the novelty electronic signal control.

**Methods.**

**Sample fabrications and circuit signal measurements.** We exploit the electric circuits by using PADs program software, where the PCB composition, stack up layout, internal layer and grounding design are suitably engineered. Here, the well-designed PCB possesses totally eight layers to arrange the intralayer and interlayer site-couplings. It is worthy to note that the ground layer should be placed in the gap between any two layers to avoid their coupling. Moreover, all PCB traces have a relatively large width (0.5mm) to reduce the parasitic inductance and the spacing between electronic devices is also large enough (1.0mm) to avert spurious inductive coupling. The SMP connectors are welded on the PCB nodes for the signal input and detection. To ensure the realization of BOs in electric circuits, both the tolerance of circuit elements and series resistance of inductors should be as low as possible. For this purpose, we use WK6500B impedance analyzer to select circuit elements with high accuracy (the disorder strength is only 1%) and low losses.

As for the measurement of BOs, we use the signal generator (NI PXI-5404) with eight output ports to act as the current source for exciting two circuit nodes related to a single lattice site with a constant amplitude and node-dependent initial phases. One output of the signal generator (the initial phase is set to 0) is directly connected to one end of the oscilloscope (Agilent Technologies Infiniivision DSO7104B) to make sure an accurate start time. The measured voltage signals are in the range from 0ms to 2ms in the time-domain, where the 0ms is defined as the time for the simultaneously signal injection and measurement.

**Acknowledgements**

This work was supported by the National key R & D Program of China under Grant No. 2017YFA0303800 and the National Natural Science Foundation of China (No.91850205 and No.61421001).

# Supplementary Materials

S1. Details for the derivation of the eigen-equation for the 2D circuit simulator with $\theta = \pi$.

S2. The influence of the value of $C_e$ on the correspondence between eigen-spectra of 2D circuit simulators and 1D two-anyon models.

S3. Numerical results of Bloch oscillations based on the 1D extended anyon-Hubbard model.

S4. Simulating the anyonic Bloch oscillation with different excitation frequencies, external forces and grounding capacitor $C_e$.

S5. The precise correspondence between time-dependent Schrödinger equation of two bosons and two pseudofermions and designed RC circuit simulators.

**S1. Details for the derivation of the eigen-equation for the 2D circuit simulator with $\theta = \pi$.**

In this part, we give a detailed derivation of circuit eigen-equation, which could be mapped to the 1D stationary Schrödinger equation of two pseudofermions. Here, each lattice site possesses two circuit nodes. In this case, the voltage and current at the site ($m, n$) should be written as $V_{(m,n)} = [V_{(m,n),1}, V_{(m,n),2}]^T$ and $I_{(m,n)} = [I_{(m,n),1}, I_{(m,n),2}]^T$. And, the voltage on the circuit node ($m, n$) is in the form of $V_{(m,n),j} e^{i\omega t}$ ($j=1$ and 2).

At first, we focus on the node pair located at the diagonal line ($n, n$). Carrying out the Kirchhoff's law on the circuit node pair ($n, n$), we get the following equation as:

$$\begin{vmatrix} I_{(n,n),1} \\ I_{(n,n),2} \end{vmatrix} = i\omega^{-1} \{ -\frac{1}{L} \begin{vmatrix} 1 & -1 \\ -1 & 1 \end{vmatrix} \begin{vmatrix} V_{(n,n),1} \\ V_{(n,n),2} \end{vmatrix} + \omega^2 C \begin{vmatrix} V_{(n,n),1} - V_{(n-1,n),1} \\ V_{(n,n),2} - V_{(n-1,n),2} \end{vmatrix} + \omega^2 C \begin{vmatrix} V_{(n,n),1} - V_{(n+1,n),1} \\ V_{(n,n),2} - V_{(n+1,n),2} \end{vmatrix}$$

$$+ \omega^2 C \begin{vmatrix} V_{(n,n),1} - V_{(n,n-1),2} \\ V_{(n,n),2} - V_{(n,n-1),1} \end{vmatrix} + \omega^2 C \begin{vmatrix} V_{(n,n),1} - V_{(n,n+1),2} \\ V_{(n,n),2} - V_{(n,n+1),1} \end{vmatrix}$$

$$+ \omega^2 C_U \begin{vmatrix} V_{(n,n),1} \\ V_{(n,n),2} \end{vmatrix} + \omega^2 (n+n) C_F \begin{vmatrix} V_{(n,n),1} \\ V_{(n,n),2} \end{vmatrix} + \omega^2 C_e \begin{vmatrix} V_{(n,n),1} \\ V_{(n,n),2} \end{vmatrix} \}, \tag{S1}$$

where $C_U$, $C_e$ and $(n+n)C_F$ are capacitances linking the node ($n, n$) to the ground. $C$ is the capacitance used for connecting circuit nodes belonging to adjacent lattice sites. $L$ is the inductor linking the circuit nodes belonging to the same lattice site.

We assume that there is no external source, so that the current flowing out of the node is zero. In this case, Eq. (S1) becomes:

$$\frac{1}{\omega^2 L}\begin{vmatrix} 1 & -1 \\ -1 & 1 \end{vmatrix}\begin{vmatrix} V_{(n,n),1} \\ V_{(n,n),2} \end{vmatrix} = [4C + C_U + C_e + (n+n)C_F]\begin{vmatrix} V_{(n,n),1} \\ V_{(n,n),2} \end{vmatrix}$$

$$-C\begin{bmatrix} 0 & 1 \\ 1 & 0 \end{bmatrix}(\begin{vmatrix} V_{(n,n-1),2} \\ V_{(n,n-1),1} \end{vmatrix} + \begin{vmatrix} V_{(n,n+1),2} \\ V_{(n,n+1),1} \end{vmatrix}) - C\begin{vmatrix} V_{(n-1,n),1} \\ V_{(n-1,n),2} \end{vmatrix} - C\begin{vmatrix} V_{(n+1,n),1} \\ V_{(n+1,n),2} \end{vmatrix}. \quad (S2)$$

Performing the diagonalization of Eq. (S2) with a unitary transformation:

$$F = \frac{1}{\sqrt{2}}\begin{bmatrix} 1 & e^{i\pi} \\ 1 & -e^{i\pi} \end{bmatrix}. \quad (S3)$$

Eq. (S2) becomes:

$$\frac{1}{\omega^2 L}\begin{vmatrix} 0 & 0 \\ 0 & 2 \end{vmatrix}\begin{vmatrix} V_{\uparrow,(n,n)} \\ V_{\downarrow,(n,n)} \end{vmatrix} = [4C + C_U + C_e + (n+n)C_F]\begin{vmatrix} V_{\uparrow,(n,n)} \\ V_{\downarrow,(n,n)} \end{vmatrix} - C\begin{bmatrix} 1 & 0 \\ 0 & e^{-i\pi} \end{bmatrix}\begin{vmatrix} V_{\uparrow,(n,n-1)} \\ V_{\downarrow,(n,n-1)} \end{vmatrix}$$

$$-C\begin{bmatrix} 1 & 0 \\ 0 & e^{i\pi} \end{bmatrix}\begin{vmatrix} V_{\uparrow,(n,n+1)} \\ V_{\downarrow,(n,n+1)} \end{vmatrix} - C\begin{vmatrix} V_{\uparrow,(n-1,n)} \\ V_{\downarrow,(n-1,n)} \end{vmatrix} - C\begin{vmatrix} V_{\uparrow,(n,n+1)} \\ V_{\downarrow,(n,n+1)} \end{vmatrix}\}. \quad (S4)$$

The new basis are $V_{\uparrow(\downarrow),(m,n)} = F[V_{(m,n),1}, V_{(m,n),2}]^T$, which are two decoupled terms acting as a pair of pseudospins $V_{\uparrow,(m,n)} = (V_{(m,n),1} + V_{(m,n),2})/\sqrt{2}$ and $V_{\downarrow,(m,n)} = (V_{(m,n),1} - V_{(m,n),2})/\sqrt{2}$. Thus, Eq. (S4) can be divided into two independent equations as:

$$0 = [4C + C_U + C_e + (n+n)C_F]V_{\uparrow,(n,n)} - C(V_{\uparrow,(n,n-1)} + V_{\uparrow,(n,n+1)} + V_{\uparrow,(n-1,n)} + V_{\uparrow,(n,n+1)}), \quad (S5)$$

$$\frac{1}{\omega^2 L/2}V_{\downarrow,(n,n)} = [4C + C_U + C_e + 2nC_F]V_{\downarrow,(n,n)} - C(e^{-i\pi}V_{\downarrow,(n,n-1)} + e^{i\pi}V_{\downarrow,(n,n+1)} + V_{\downarrow,(n-1,n)} + V_{\downarrow,(n,n+1)}).(S6)$$

Based on similar derivations, we can write the eigen-equation at any circuit node (*m*, *n*). As for the case of *n*=*m*-1, we have,

$$\frac{1}{\omega^2 L/2}V_{\downarrow,(n,n-1)} = [4C + C_U + C_e + 2nC_F]V_{\downarrow,(n,n-1)} - C(V_{\downarrow,(n,n-2)} + e^{i\pi}V_{\downarrow,(n,n)} + V_{\downarrow,(n-1,n-1)} + V_{\downarrow,(n+1,n-1)}).(S7)$$

As for the case of *n*=*m*+1, we have,

$$\frac{1}{\omega^2 L/2}V_{\downarrow,(n,n+1)} = [4C + C_U + C_e + 2nC_F]V_{\downarrow,(n,n+1)} - C(e^{-i\pi}V_{\downarrow,(n,n)} + V_{\downarrow,(n,n+2)} + V_{\downarrow,(n-1,n+1)} + V_{\downarrow,(n+1,n+1)}). \quad (S8)$$

As for the case of $|n-m| > 1$, we have,

$$\frac{1}{\omega^2 L/2} V_{\downarrow,(m,n)} = [4C + C_U + C_e + 2nC_F]V_{\downarrow,(m,n)} - C(V_{\downarrow,(m,n-1)} + V_{\downarrow,(m,n+1)} + V_{\downarrow,(m+1,n)} + V_{\downarrow,(m-1,n)}). \quad (S9)$$

Combing Eqs. (S6)-(S9), the eigen-equation of the designed circuit simulator is described by:

$$(f_0^2/f^2 - 4 - C_e/C)V_{\downarrow,mn} = -e^{-i\pi(\delta_{m,n}+\delta_{m,n+1})}V_{\downarrow,m(n+1)} - e^{i\pi(\delta_{m,n}+\delta_{m+1,n})}V_{\downarrow,m(n-1)}$$

$$-V_{\downarrow,(m+1)n} - V_{\downarrow,(m-1)n} + (C_U/C)V_{\downarrow,mn} + (m+n)(C_F/C)V_{\downarrow,mn}. \quad (S10)$$

We provide the following identification of tight-binding parameters in terms of circuit elements as:

$$J = 1, U = \frac{C_U}{C}, \quad F = \frac{C_F}{C}, \quad \varepsilon = \frac{f_0^2}{f^2} - 4 - \frac{C_e}{C}, \quad f_0 = \frac{1}{2\pi\sqrt{CL/2}}, \quad (S11)$$

where $J$, $F$, $U$ and $\varepsilon$ correspond to the strength of the particle hopping, external forcing, on-site interaction and the eigen-energy of two anyons. In this case, Eq. (S10) becomes:

$$\varepsilon c_{mn} = -J[e^{i\theta(\delta_{m,n}+\delta_{m+1,n})}c_{m(n-1)} + e^{-i\theta(\delta_{m,n}+\delta_{m-1,n})}c_{m(n+1)} + c_{(m-1)n} + c_{(m+1)n}]$$

$$+U\delta_{mn}c_{mn} + F(m+n)c_{mn} \quad (S12)$$

with $c_{mn}$ corresponding to $V_{\downarrow,(m,n)}$. It is noted that Eq. (S12) is consistent with the eigen-equation of $c_{mn}$ for the 1D pseudofermions (Eq. (4) in the main text).

**S2. The influence of the value of $C_e$ on the correspondence between eigen-spectra of 2D circuit simulators and 1D two-anyon models.**

It is known that the appearance of Bloch oscillations depends on the equally spaced eigen-spectrum of two bosons and two pseudofermions, and the periods are determined by the associated energy level spacings. While, due to the nonlinear relationship between the eigen-frequency of circuit simulator and the eigen-energy of two anyons $f = f_0/(\varepsilon + 4 + C_e/C)^{1/2}$, the distribution of eigen-spectrum for the circuit simulator should not be equally spaced.

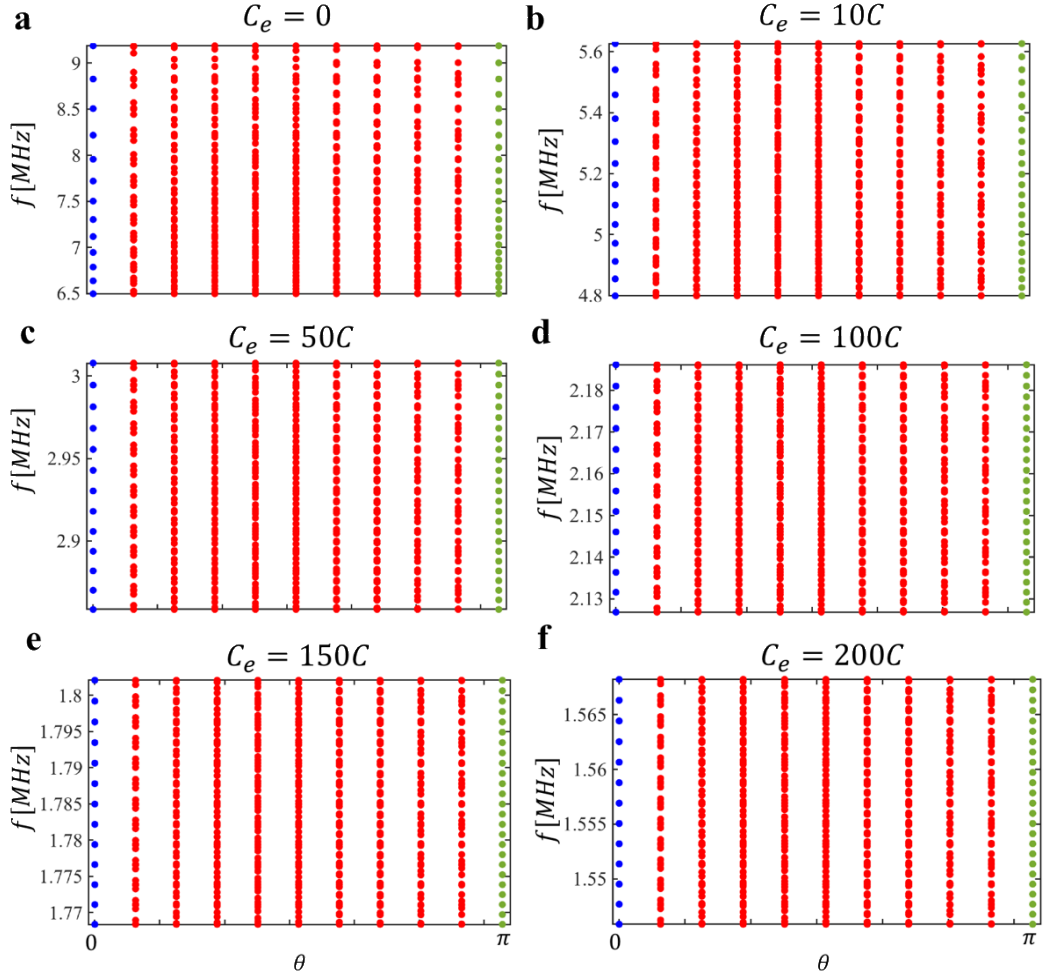

**Fig. S1.** (a)-(f) plot eigen-frequencies of designed circuit simulators as a function of the statistical angle $\theta$ with $C_e = 0$, $C_e = 10C$, $C_e = 50C$, $C_e = 100C$, $C_e = 150C$, and $C_e = 200C$, respectively.

In this part, we show that such a deviation could become negligible with a large value of $C_e$ used in our circuit design. As shown in Figs. S1(a)-S1(f), we plot eigen-frequencies of designed circuit simulators as a function of the statistical angle $\theta$ with $C_e = 0$, $C_e = 10C$, $C_e = 50C$, $C_e = 100C$, $C_e = 150C$, and $C_e = 200C$, respectively. It is clearly shown that the eigen-spectrum with $C_e = 0$ is not equally spaced for the circuit simulator with $\theta = 0$ and $\theta = \pi$, where the frequency spacing is getting increased with the increase of eigen-frequencies. By increasing the value of $C_e$, the difference of frequency spacings belonging to the higher and lower eigen-frequencies ranges is decreased. In our design, we set $C_e = 200C$. In this case, we can see that nearly equal-spaced eigen-spectra of circuit simulators with $\theta = 0$ ($\Delta f_B \approx 1862.57 Hz$) and $\theta = \pi$ ($\Delta f_f \approx 931.28 Hz$) appear. With such a good correspondence between the eigen-spectrum of designed 2D circuit and the 1D two-anyon model, the behavior of the quantum statistics dominated Bloch oscillation can be effectively simulated by the designed circuit simulator.

**S3. Numerical results of Bloch oscillations based on the 1D extended anyon-Hubbard model.**

In this part, we give numerical results of BOs described by the 1D extended anyon-Hubbard model with $N=23$. The evolution equations for the probability amplitude $c_{mn}$ can be obtained by substituting Eqs. (1) and (3) (in the main text) into the time-dependent Schrödinger equation $H|\psi> = i\frac{\partial}{\partial t}|\psi>$. In this case, we get

$$i\partial_t c_{mn} = -J[e^{i\theta(\delta_{m,n}+\delta_{m+1,n})}c_{m(n-1)} + e^{-i\theta(\delta_{m,n}+\delta_{m-1,n})}c_{m(n+1)} + c_{(m-1)n} + c_{(m+1)n}]$$

$$+ U\delta_{mn}c_{mn} + F(m+n)c_{mn} . \tag{S13}$$

To observe the BO of bosons and pseudofermions, the external excitation is set as:

$$c_{12,12}(t) = e^{i\varepsilon t} \tag{S14}$$

Here, other parameters are set as $\varepsilon = 20$, $J=1$, $U=0$ and $F=0.5$, respectively.

As shown in Figs. S2a and S2b, we calculate the evolution of $|c_{mn}(t)|^2$ with $\theta = 0$ and $\theta = \pi$, respectively. Moreover, Figs. S2c and S2d display the evolution of $|c_{12,12}(t)|$ with $\theta = 0$ and $\theta = \pi$, respectively. It is clearly shown that periodic breathing dynamics of both bosons and pseudofermions appear, and the oscillation period of the two bosons is almost twice of that of two pseudofermions.

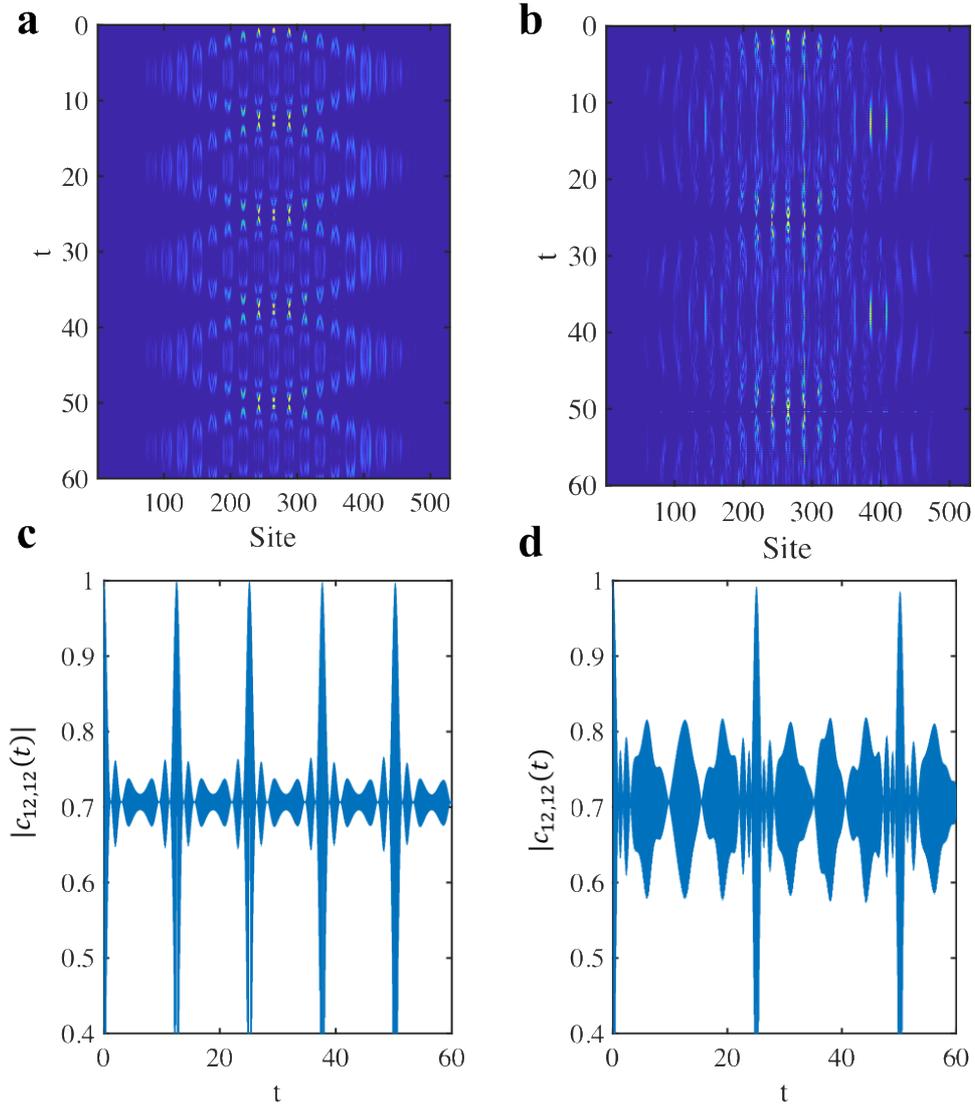

**Fig. S2.** (a) and (b) The evolution of $|c_{mn}(t)|^2$ with $\theta = 0$ and $\theta = \pi$ in the absence of particle interactions, respectively. (c) and (d) display the evolution of $|c_{12,12}(t)|$ with $\theta = 0$ and $\theta = \pi$. Here, parameters are set as $J=1$, $U=0$ and $F=0.5$, respectively.

Then, we focus on the anyonic BOs with $F=0.3$. Figs. S3a and S3b present the calculated evolutions of $|c_{mn}(t)|^2$ with $\theta = 0$ and $\theta = \pi$, respectively. And, Figs. S3c and S3d display the associated evolution of $|c_{12,12}(t)|$ with $\theta = 0$ and $\theta = \pi$. It is clearly shown the larger the external force is, the larger the oscillation period and amplitude become. And, it is noted that the BO frequency related to a pair of pseudofermions ($\theta = \pi$) is always half of that for two bosons ($\theta = 0$).

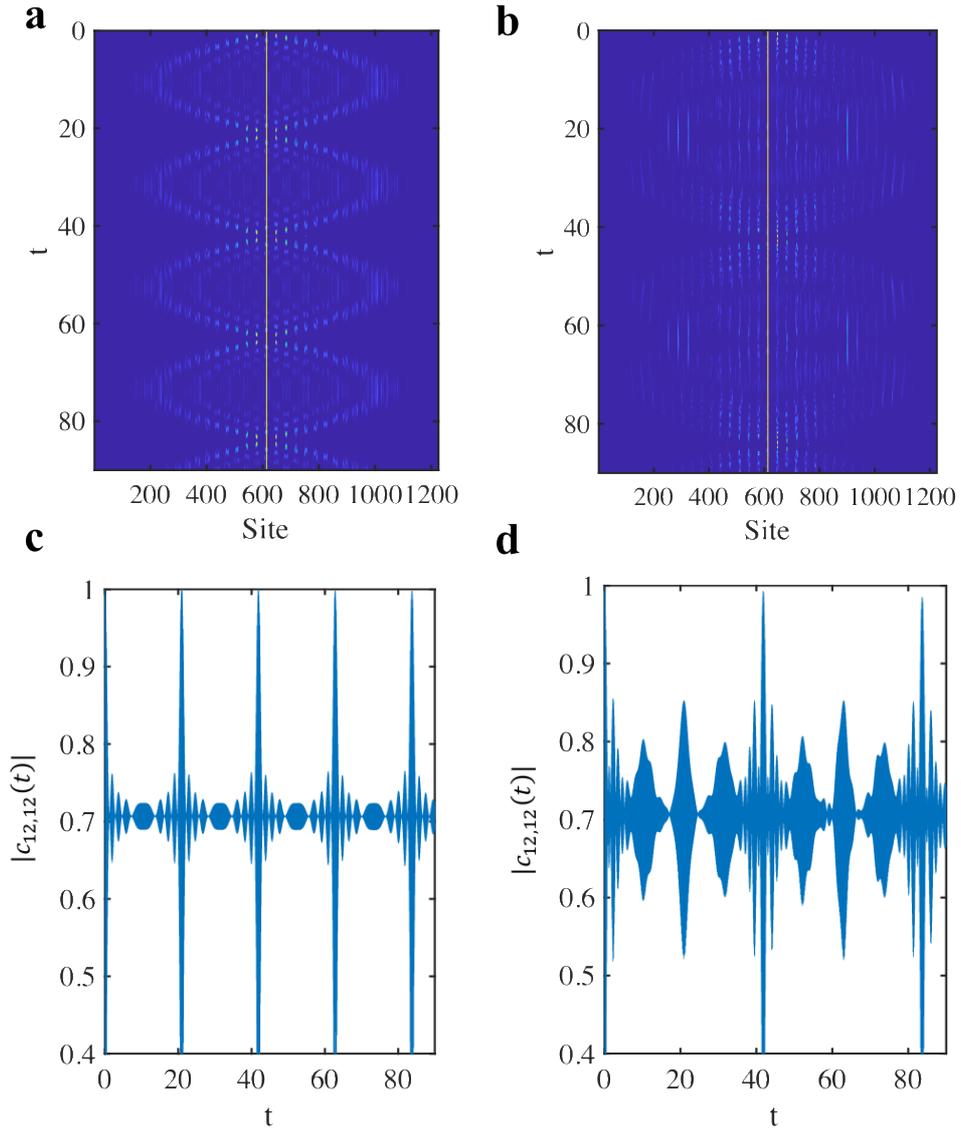

**Fig. S3.** (a) and (b) The evolution of $|c_{mn}(t)|^2$ with $\theta = 0$ and $\theta = \pi$ in the absence of particle interactions, respectively. (c) and (d) display the evolution of the particle density function of $|c_{12,12}(t)|$ with $\theta = 0$ and $\theta = \pi$. Here, parameters are set as $J$=1, $U$=0 and $F$=0.3, respectively.

**S4. Simulating the anyonic Bloch oscillation with different excitation frequencies, external forces and grounding capacitor $C_e$.**

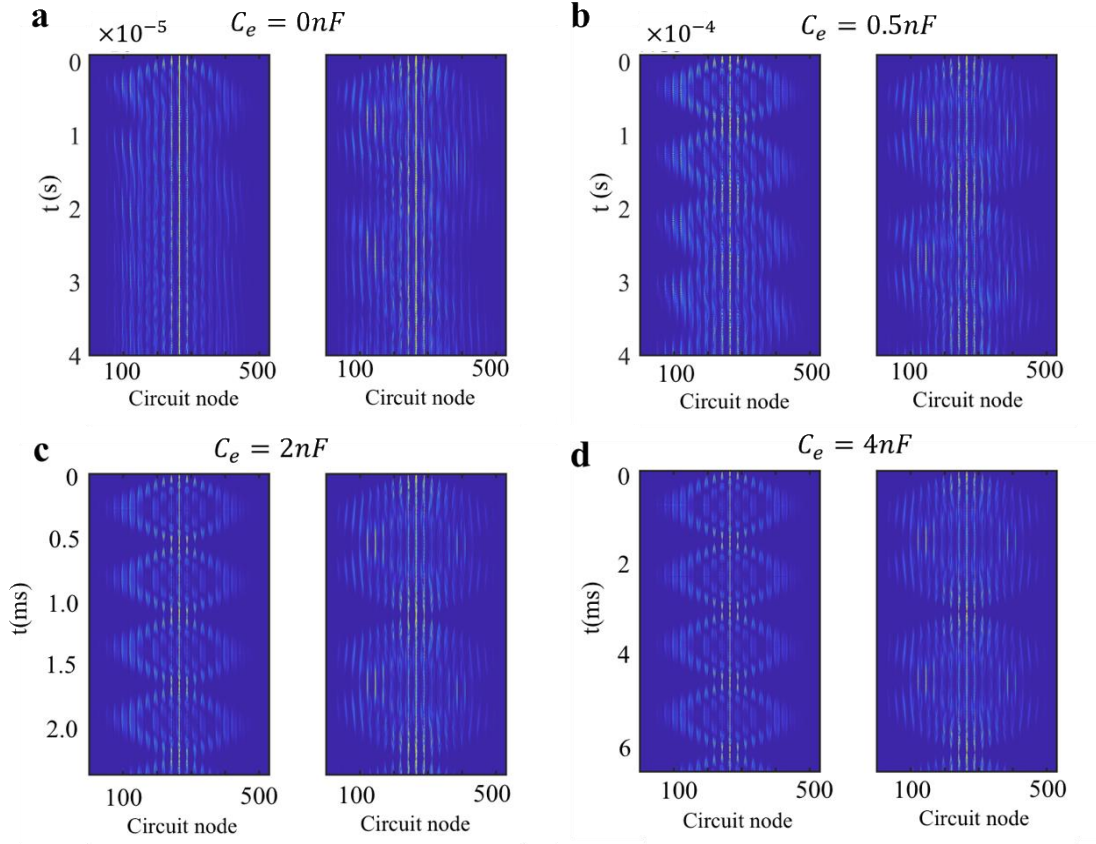

**Fig. S4.** (a)-(d) The time-dependent evolution of pseudospin $|V_{\downarrow,[m,n]}(t)|^2$ at each node in the circuit simulator with $C_e = 0$, $C_e = 0.5nF$, $C_e = 2nF$ and $C_e = 4nF$.

At first, we perform circuit simulations of BOs with different values of $C_e$. As shown in Figs. 4a-4d, we calculate the time-dependent evolution of pseudospin $|V_{\downarrow,[m,n]}(t)|^2$ at each node in the 2D circuit simulator (the left chart with $\theta = 0$ and the right chart with $\theta = \pi$) with $C_e = 0$, $C_e = 0.5nF$, $C_e = 2nF$ and $C_e = 4nF$, respectively, where the associated excitation frequencies are set as 9.19MHz, 3.01MHz, 1.56MHz, and 1.117MHz. Other parameters are the same to that used in Fig. 2. We can see that the larger the value of $C_e$ is, the more ideal BOs appear. This is due to the fact that the nearly perfect eigen-spectrum with equal spacings could only be realized with an extremely large value of $C_e$, as shown in Fig. S1.

Then, we will simulate anyonic Bloch oscillations with a different external force by our designed electric circuits, that is $C_F = 3pF$. Before circuit simulations, we calculate the evolution of two-anyon eigen-energies as a function of $\theta$ with $J=1$ and $F=0.3$, as shown in Fig. S5a. And, the eigen-frequencies of designed circuit simulators with $C_e = 0$, $C_e = 50C$, and $C_e = 200C$ are shown in Figs. S5(b)-S5(d). It is shown that the eigen-spectrum of the circuit simulator is consistent with that of two anyons with a large value of $C_e$. In particular, we have $\Delta f_B \approx 1130 Hz$ and $\Delta f_f = 565 Hz$. Next, we calculate the time-dependent evolution of pseudospin $|V_{\downarrow,[m,n]}(t)|^2$ at each node in the 2D circuit simulator ($C_e = 200C$ and $C_F = 0.3C$) with $\theta = 0$ and $\theta = \pi$, as shown in Fig. S6a and S6b. Here, the excitation frequency is set as 1.511MHz, and the voltage-pseudospin is excited by setting the input signal as $[V_{(12,12),1} = V_0, V_{(12,12),2} = -V_0]$ with $V_0 = 1V$. It is shown that the absolute value of pseudospin displays the periodic breathing dynamics for both conditions. Moreover, we note that the oscillation periods of bosonic circuits ($T_B \approx \frac{1}{\Delta f_B} = 0.885ms$) is nearly the half of the oscillation period for two pseudofermions ($T_f \approx \frac{1}{\Delta f_f} = 1.77ms$). Comparing to the results with $C_F = 0.5C$, we find that the smaller the external force is, the larger the oscillation period and amplitude become.

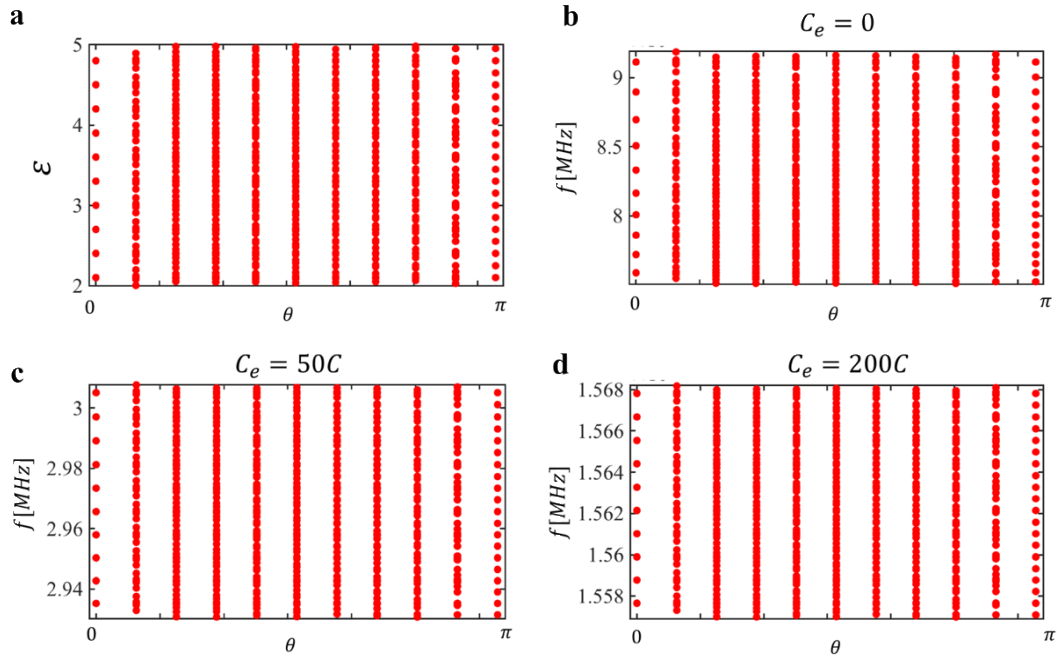

**Fig. S5.** (a). The evolution of two-anyon eigen-energies as a function of $\theta$ with $J=1$ and $F=0.3$. (b)-(d) The eigen-frequencies of circuit simulators ($C_F = 0.3C$) with $C_e = 0$, $C_e = 50C$, and $C_e = 200C$.

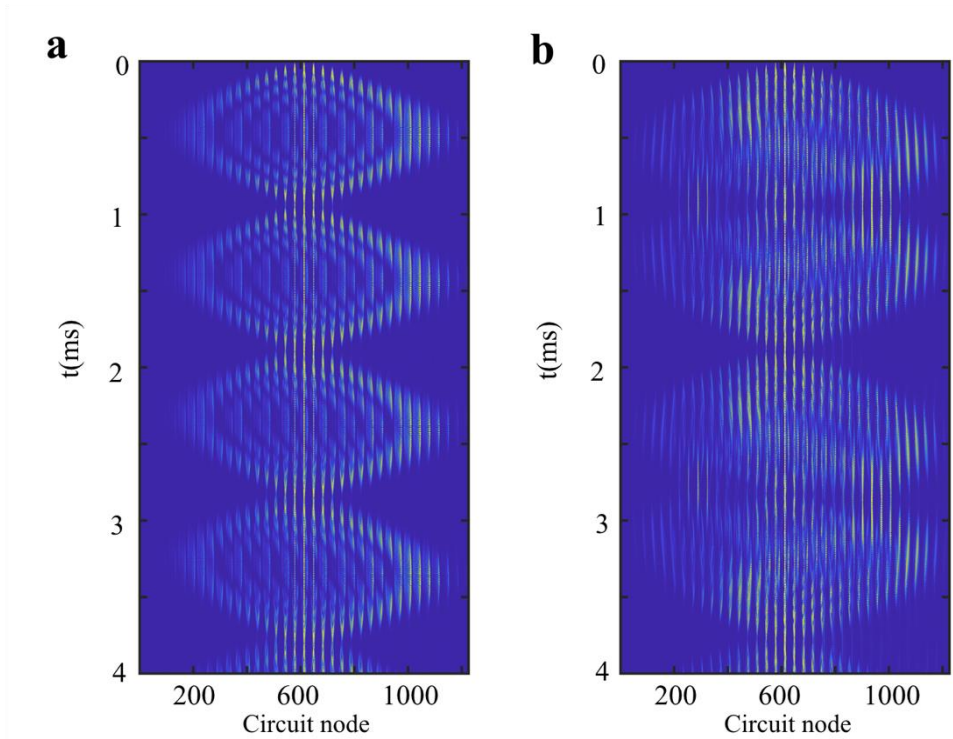

**Fig. S6.** The time-dependent evolution of pseudospin $|V_{\downarrow,[m,n]}(t)|^2$ at each node in the circuit simulator ($C_e = 200C$ and $C_F = 0.3C$) with $\theta = 0$ for (a), and $\theta = \pi$ for (b).

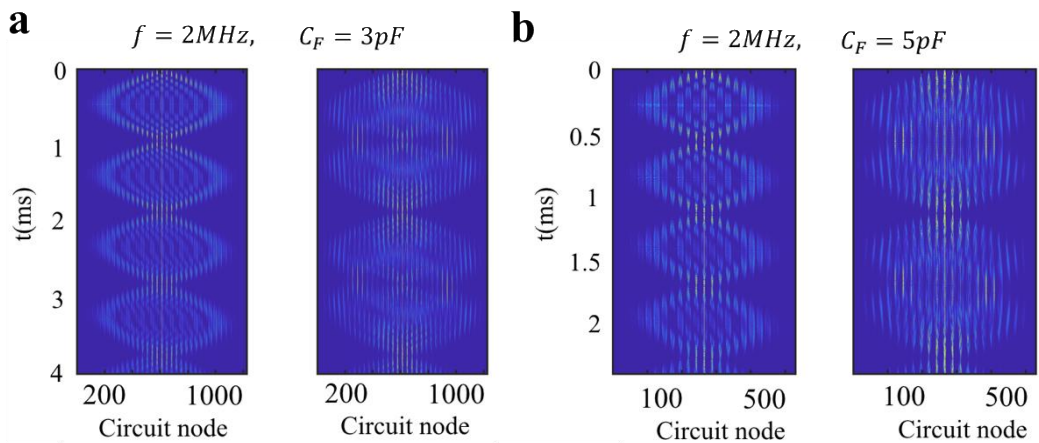

**Fig. S7.** The time-dependent evolution of pseudospin $|V_{\downarrow,[m,n]}(t)|^2$ at each node in the circuit simulator

with $C_F = 0.3C$ for (a), and $C_F = 0.5C$ for (b). The associated excitation frequency is 2MHz.

At last, we perform circuit simulations of BOs under high excitation frequencies. The time-dependent evolution of pseudospin $|V_{\downarrow,[m,n]}(t)|^2$ at each node in the 2D circuit simulator with $C_F = 0.3C$ and $C_F = 0.5C$ are shown in Figs. S7a and S7b, where the associated excitation frequency is 2MHz. Other parameters are set as $C_e = 2nF$, $C = 10pF$, and $L = 10uH$. Comparing to the corresponding results with lower excitation frequencies (Fig. 2 for $C_F = 0.5C$ and Fig. S6 for $C_F = 0.3C$), we find that the more symmetric BO could be realized with a higher excitation frequency.

**S5. The precise correspondence between time-dependent Schrödinger equation of two bosons and two pseudofermions and designed RC circuit simulators.**

It is worthy to note that the stationary eigen-equation of our designed *LC* circuit is consistent with the stationary Schrödinger equation of the 1D anyon-Hubbard model with two anyons. As for the time-dependent evolution equation, the voltage of *LC* circuit follows second-order time differential, which is different from the first-order time differential of quantum wave functions. In this part, we will design another kind of electric circuit based on resistances and capacitances to precisely match the time-dependent Schrödinger equation of two anyons with $\theta = 0$ and $\theta = \pi$.

The designed *RC* circuit simulator with $\theta = 0$ is plotted in Fig. S8a. Here, the associated 1D lattice length is *N*. We note that the designed circuit simulator contain $2N^2$ nodes, where the row (column) of $N^2$ nodes are labeled by *r*=(1, 1),…,(*N*, *N*) [*c*=(*N*+1,*N*+1),…,(2*N*, 2*N*)]. The voltages of totally $N^2$ circuit node in the top-row (left-column) correspond to the (copy of) probability amplitudes of two anyons with $\theta = 0$ in the 1D lattice of *N* sites. Specifically, the probability amplitude of two-boson states $c_{mn}$ is mapped to the voltage signal on the circuit node (*m*, *n*) as $V_{m,n}$. Each node is connected to an external DC through a switch to apply an initial voltage signal. Two nodes (with one from the row and the other from the column) are connected by a suitably designed negative impedance converters with current inversion (INICs), named as $R_{rc}$, to realize the hopping, on-site interactions and external forcing. Specially, the designed INIDs for realizing the particle hoppling rate ($R_{rc} = R_J$) and the external force ($R_{rc} = R_F/(m+n)$) are enclosed by yellow and red blocks, respectively. Here, we set the on-site interaction as zero. As for the grounding, the green (blue) circuit node in the row (column) is grounded with a constant capacitor *C* and an INIC (normal resistor) with the effective resistance being $R_{r0}$ ($R_{c0}$). In this case, the effective hoppling rate between node *r*=(*m*, *n*) and *c*=(*m'*, *n'*) is $J = \frac{1}{CR_J}$, where the node locations should satisfy the relation of $n' = n \pm 1 + N$ and $m' = m + N$ or $n' = n + N$ and $m' = m \pm 1 + N$. The external force could be mapped to the position-dependent grounding $R_F/(m+n)$ with *r*=(*m,n*) and *c*=(*m+N,n+N*). In this case, the effective external force is $F = \frac{1}{CR_F}$. The detailed node connections are plotted in the right-bottom part of Fig. S8a.

By switching off all switches at the same time (applying an initial state), the evolution of voltage at each circuit node can be derived based on the time-dependent Kirchhoff's equation as:

$$C\frac{dV_r}{dt} - \frac{V_r}{R_{r0}} = \sum_c \frac{V_c - V_r}{R_{rc}},$$

$$C\frac{dV_c}{dt} + \frac{V_c}{R_{c0}} = \sum_r \frac{V_r - V_c}{-R_{rc}} \quad (S15)$$

with $V_r$ ($V_c$) being the voltage at the circuit node in the row (column). The summation is limited to the connected circuit nodes. Defining the voltages at all circuit nodes as $|V(t)\rangle = [V_{(1,1)}(t), \ldots, V_{(N,N)}(t), V_{(N+1,N+1)}(t), \ldots, V_{(2N,2N)}(t)]$, Eq. (S15) could be expressed in the matrix form as $i\partial_t |V(t)\rangle = \Xi |V(t)\rangle$, where the off-diagonal components of circuit Hamiltonian $\Xi$ are $\Xi_{rc} = i\frac{1}{CR_{rc}}$ and $\Xi_{cr} = -i\frac{1}{CR_{rc}}$, and the diagonal components are given by $\Xi_{rr} = i\frac{1}{C}(\frac{1}{R_{r0}} - \sum_c \frac{1}{R_{rc}})$ and

$\Xi_{cc} = i\frac{1}{C}(-\frac{1}{R_{c0}} + \sum_r \frac{1}{R_{rc}})$. By appropriately setting the grounding INICs as $\frac{1}{R_{ro(i)}} = \sum_k \frac{1}{R_{ik}}$ and the grounding resistances as $\frac{1}{R_{co(i)}} = \sum_k \frac{1}{R_{ik}}$, the circuit Hamiltonian can be expressed as:

$$\Xi = i\begin{vmatrix} 0 & -\Pi \\ \Pi & 0 \end{vmatrix} \tag{S16}$$

with $\Pi$ being a $N \times N$ matrix. In this case, when the nodes connecting and grounding resistances are suitably applied, the form of $N \times N$ matrix $\Pi$ can be the same to the Hamiltonian of the 1D two-boson model. In this case, the voltage evolution in the designed RC circuit could be the same to the probability amplitude of two bosons.

Based on the similar method, the RC circuit related to two pseudofermions $\theta = \pi$ could also be designed. Fig. S8b presents the corresponding connection pattern at different circuit nodes. Comparing to the circuit for two bosons, the only difference is that there are a few of effective hopping rates sustaining a phase $e^{i\pi}$. This could be easily fulfilled by reversing the biased voltage of the associated grounding and connecting INICs.

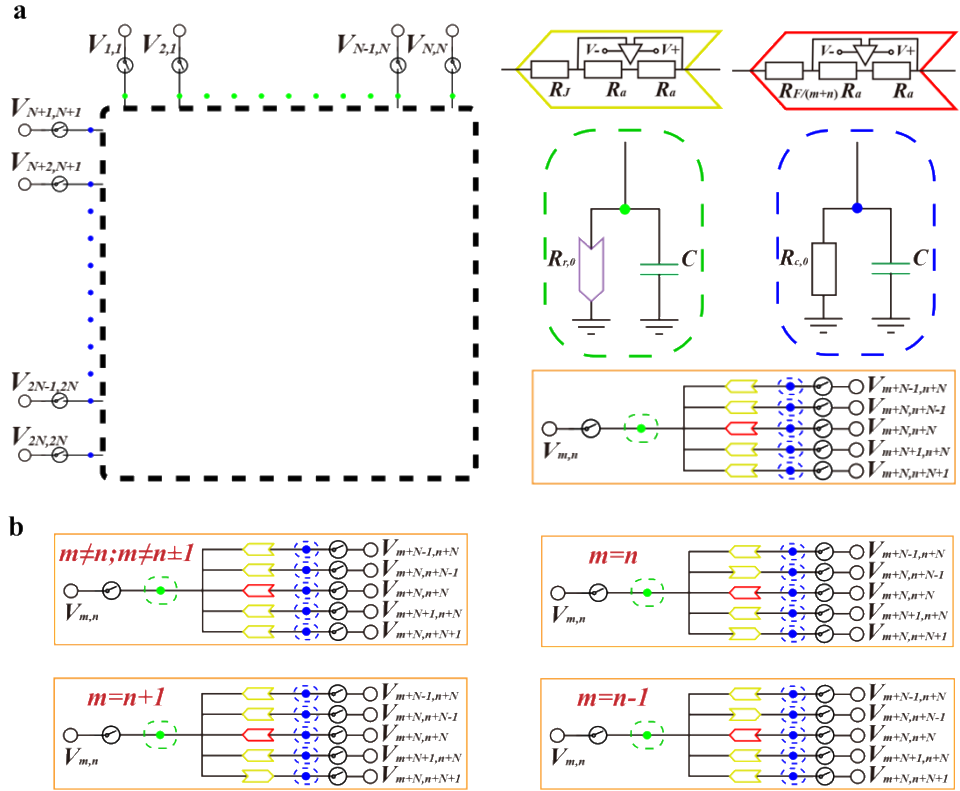

**Fig. S8.** (a) and (b) The designed *RC* circuit for simulating Bloch oscillations of two bosons and two pseudofermions.

Then, we use the designed *RC* circuit to simulate the BOs of two bosons and two pseudofermions. Other parameters are set as *N*=35, *C*=1uF, $R_J = 1000\Omega$, $R_F = 2000\Omega$ and $R_a = 100\Omega$. And, the initial voltage distribution is set as: $V_{mn}(t = 0) = V_0\delta_{12,12}$. As shown in Fig. S9a and S9b, we present the calculated evolution of the signal $|V_{\downarrow,[m,n]}(t)|^2$ in the circuit simulators for two bosons and two pseudofermions, respectively. And, the corresponding time-dependent evolutions of $|V_{\downarrow,[12,12]}|^2$ are presented in Figs. S9c and S9d.

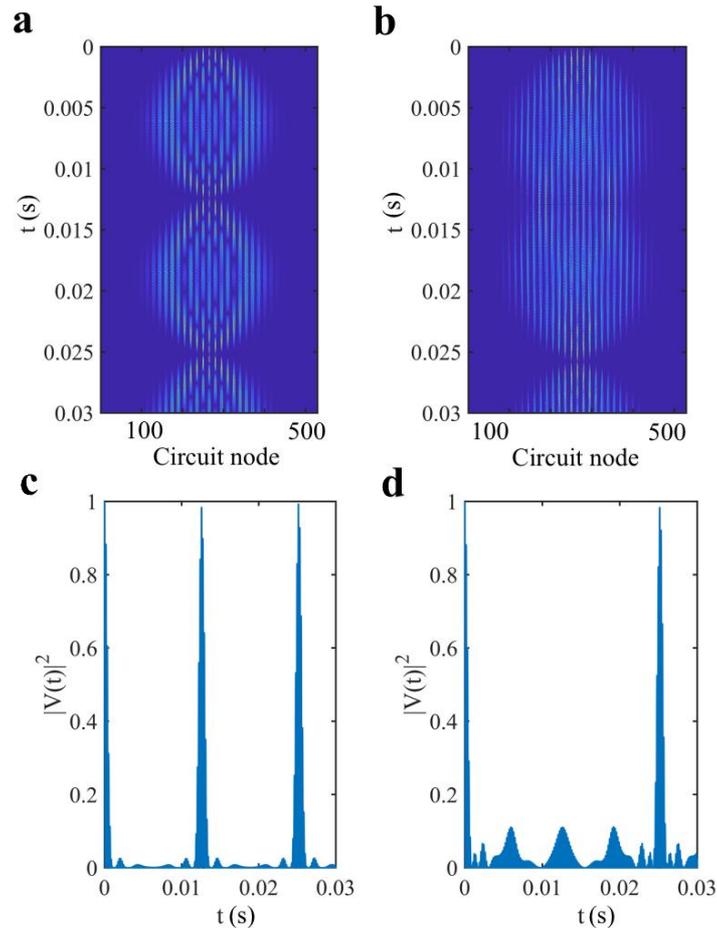

**Fig. S9.** (a) and (b) The evolution of the signal $|V_{\downarrow,[m,n]}(t)|^2$ in the circuit simulators for two bosons

and two pseudofermions, respectively. (c) and (d) The time-dependent evolutions of $|V_{\downarrow,[12,12]}|^2$ for two bosons and two pseudofermions.

For comparations, we also calculate the evolution of $|c_{mn}(t)|^2$ of two bosons and two pseudofermions in the 1D anyon-Hubbard model, as shown in Figs. S10a and S10b. The associated parameters are set as $J=1$, $F=0.5$ and $C_{mn}(t=0) = \delta_{12,12}$. And, the corresponding time-dependent evolutions of $|c_{12,12}|^2$ are presented in Figs. S10c and S10d. We note that a good agreement for the time-dependent evolution of voltages and probability amplitude is obtained. In particular, it is clearly shown that the oscillation period in the two-boson simulator is twice of that in the two-pseudofermion simulator, that is consistent with the theoretical prediction.

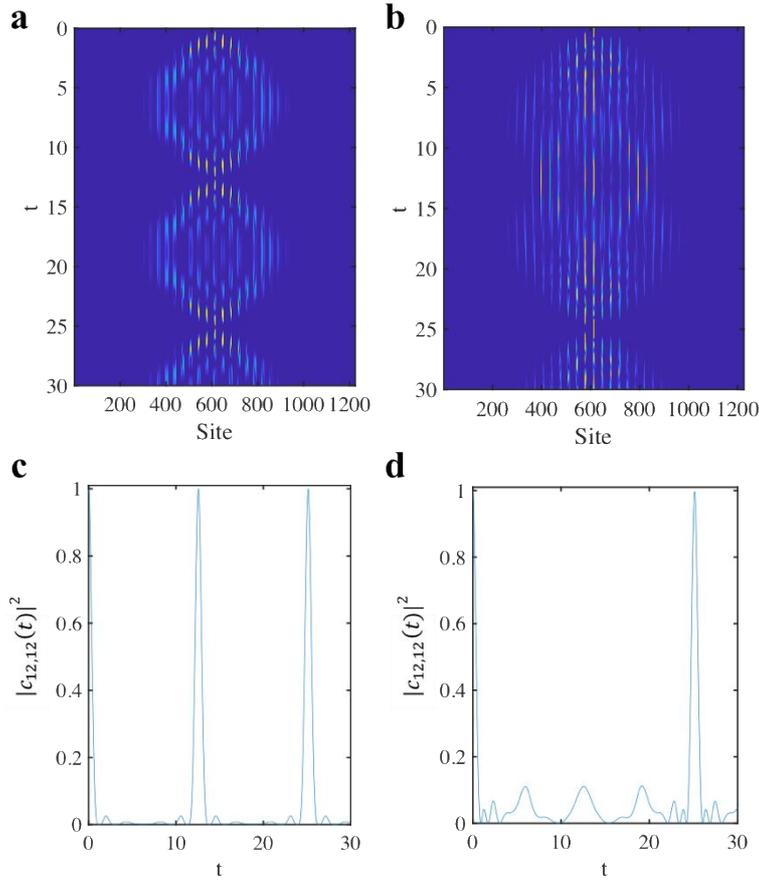

**Fig. S10.** (a) and (b) The evolution of $|c_{mn}(t)|^2$ for two bosons and two pseudofermions in the 1D

anyon-Hubbard model. (c) and (d) The time-dependent evolutions of $|c_{12,12}|^2$ for two bosons and two pseudofermions.